\documentclass{article}

\usepackage{PRIMEarxiv}
\usepackage{graphics}

\usepackage[utf8]{inputenc} % allow utf-8 input
\usepackage[T1]{fontenc}    % use 8-bit T1 fonts
\usepackage{hyperref}       % hyperlinks
\usepackage{url}            % simple URL typesetting
\usepackage{booktabs}       % professional-quality tables
\usepackage{amsfonts}       % blackboard math symbols
\usepackage{nicefrac}       % compact symbols for 1/2, etc.
\usepackage{microtype}      % microtypography
\usepackage{lipsum}
\usepackage{fancyhdr}       % header
\usepackage{graphicx}       % graphics
\graphicspath{{media/}}     % organize your images and other figures under media/ folder

\title{Skin Disease Diagnosis Using Image Analysis and Natural Language Processing
}

\author{
  Martin Chileshe \\
  Computer Science Department\\
  The University of Zambia  \\
  Lusaka\\
  \texttt{mc1a14@cs.unza.zm} \\
   \And
  Mayumbo Nyirenda \\
  Computer Science Department \\
  The University of Zambia \\
  Lusaka\\
  \texttt{mayumbo@gmail.com} \\
}

\begin{document}
\maketitle

\begin{abstract}

In Zambia, there is a serious shortage of medical staff where each practitioner attends to about 17000 patients in a given district while still, other patients travel over 10 km to access the basic medical services. In this research, we implement a deep learning model that can perform the clinical diagnosis process. The study will prove whether image analysis is capable of performing clinical diagnosis. It will also enable us to understand if we can use image analysis to lessen the workload on medical practitioners by delegating some tasks to an AI. The success of this study has the potential to increase the accessibility of medical services to Zambians, which is one of the national goals of Vision 2030 \cite{1}.

\end{abstract}

% keywords can be removed

\section{INTRODUCTION}

In medicine, there is a process called Clinical Diagnosis, which identifies a disease, condition, or injury based on the signs and symptoms a patient is having and the patient's health history and physical exam. Further testing, such as blood tests, imaging tests, and biopsies, may then follow \cite{2}. Our research targets those conditions that require no further tests after a physical examination.

Image analysis involves processing an image into fundamental components to extract meaningful information \cite{3} while Natural language processing (NLP) refers to the branch of computer science—and more specifically, the branch of artificial intelligence or AI—concerned with giving computers the ability to understand text and spoken words in much the same way human beings can \cite{4}. An indicator of the practicality of this is the level of accuracy that we will be able to achieve. Our goal is to get accuracy higher or equal to 50\%. Once we have achieved this, we shall combine it with natural language processing and use the two in a mobile application, which should perform the clinical diagnosis process on patients.

In the next section, we will briefly give an overview of the progress, advancements and gaps in image analysis, natural language processing and clinical diagnosis \ref{sec:litraturesurvey} Literature Survey. We will then proceed to \ref{sec:problem_definition} Problem Definition where we present the problem that we aim to solve then we shall proceed to \ref{sec:methodology_approach} Methodology/Approach where we indicate the steps and methods used to conduct our research. We will then go to \ref{sec:resuts_and_discussion} Results and Discussion where we show our findings as well as our thoughts on these findings. We will then \ref{sec:conclusion} Conclude our research and give our recommendations on what we think \ref{sec:future_scope} Future Scope may cover and finally provide all our \ref{sec:references} References.

\section{Literature Survey}
\label{sec:litraturesurvey}

Doctors use Skin Examination to identify skin disorders. The examination may include examination of the scalp, nails and mucus membranes while other times they may use a hand-held lens or a dermoscopy to see areas of concern. Characters that may give clues to the infection include size, shape, colour and the location of the abnormality. If looking at the skin does not provide enough information, doctors may then use other methods like a biopsy, scrapings, culture, wood light (blacklight), tzanck testing, diascopy and other skin tests \cite{5}.

\begin{figure}
  \centering
  \includegraphics[width=.6\textwidth]{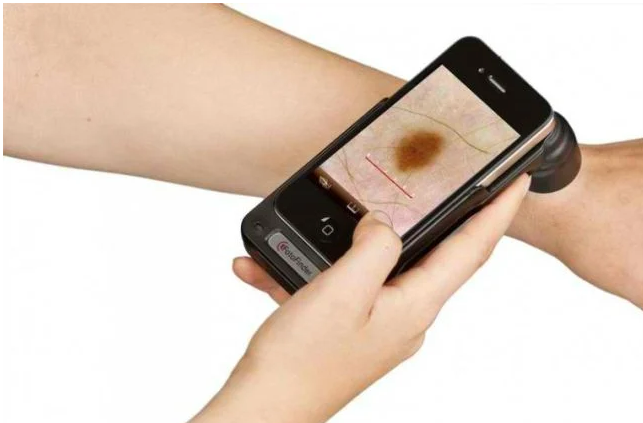}
  \caption{Analysing the Skin with a Dermascope.}
  \label{fig:dermascope}
\end{figure}

\begin{figure}
  \centering
  \includegraphics[width=.6\textwidth]{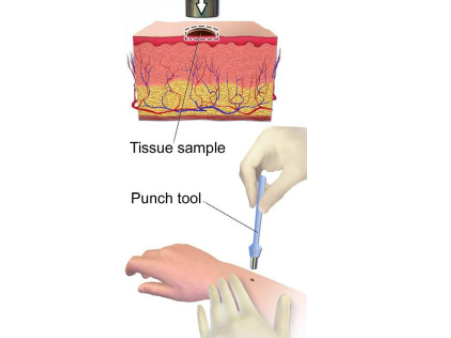}
  \caption{Punch Biopsy. To remove a small piece of skin for examination under a microscope, doctors may use a round cutting instrument. This procedure is what is known as a punch biopsy.}
  \label{fig:punch_biopsy}
\end{figure}

Image analysis has gained rapid adoption in modern medicine especially in computer-aided diagnostic approaches to provide possible disease mechanisms \cite{6} We have also seen its usage in images of MRI and CT scans as well as the detection of brain tumours. While the mentioned fields use Convolutional Neural Networks \cite{7}, ECG based predictions rely on recurrent neural networks, which are all advanced techniques in image analysis. According to Abhishek Sengupta and Priyanka Narad we expect to see more successful applications of AI-Based in the next 5-10 \cite{8} years. However, the research listed above did not target skin diagnosis but they did review a lot about image analysis in medicine.
It is also worth mentioning that the successful usage of image analysis requires huge volumes of related image data for training and evaluation which is almost impossible to obtain but thanks to modern emerging techniques like the use of synthetic data \cite{9}. A technique that we hope to use in this research.
Mobile applications have seen wide adoption in helping people manage their health and wellness to promote healthy living and gain access to useful medical information. According to industry estimates in 2017, 325,000-health care applications were available for download \cite{10}.
Some notable mobile applications include HnG a mobile app that displays and delivers a wide range of prescription medicine, medisearch, virtual doctor \cite{11}. IBM Watson for Oncology is an assistant that is being used to help provide individualized treatments for cancer patients in China. According to a study conducted by Fang-wen Zou, Babylon showed 78.2\% concordance with real-world clinical analysis. Although this is not good enough to replace humans, it does sufficed to work as a decision support system \cite{12}. Babylon is yet another tool that provides remote consultations with doctors and health care professionals via text and video messaging. Adam Baker also concluded that Babylon gives more accurate results as compared to human doctors \cite{13}. Medwhat provides medical diagnosis by referencing its large pool of expert diagnosis tips provided by doctors, engineers and researchers \cite{14}.

In a study by Petter Bae Brandtzaeg, we learn that people love Instant feedback, others feel more comfortable reviewing personal details to an AI than to a human. This shows great potential for a new paradigm that can change how Zambians receive medical services \cite{15}. In another study, Tara Qian Suna, and Rony Medagliab IBM Watson concludes that people from different cultures and societies have benefited differently from AI \cite{16}.
As indicated by the above review, we see that there has not been much research targeting skin infections in Zambia. The above It is also important to consider the local regulations provided by ZAMRA and HPCZ.

\section{Problem Definition}
\label{sec:problem_definition}

Zambia lacks adequate medical staff as reported by the World Health Organization (WHO) in 2017.  It has less than half the WHO recommended HRH workforce in all categories of the Health sector. 1:17000 Physicians, 1:13000 Nurses and Midwives; 80\% of which are in the public sector \cite{17}.
\begin{figure}[!ht]
  \centering
  \includegraphics[width=.4\textwidth]{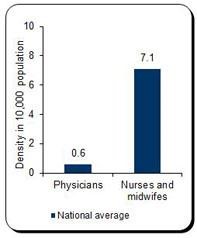}
  \caption{Patient to physician ratios}
  \label{fig:patient_ratios}
\end{figure}

\newpage
\section{Methodology/Approach}
\label{sec:methodology_approach}
\subsection{Purposive Sampling}

The first step is to collect skin related information from specialist doctors/dermatologists. A process also known as purposive sampling. This data will review what kind of skin infections have distinct features, those that are very similar to each other and those that do not always require additional diagnostic physical/clinical examination. Below is a summary of the data that was collected using a survey:

\begin{table}[!ht]
 \caption{Common Diseases}
  \centering
  \begin{tabular}{lll}
    \toprule
    Name     & Requires a lab/biopsy (yes|no) & Similar to\\
    \midrule
    Mycosis & No & Mycoses\\
    Urticaria & No & Eczema,
Angioedema
\\
    Steven-johnson syndrome & No & \\
    Kaposi’s sarcoma & No & \\
    Acne vulgaris & No & Folliculitis\\
    Eczema & No & Acne, psoriasis\\
    Photodermatitis & No &\\
    Alopecia areata & No &\\
    Vitiligo & No \\
    Dermatophytes & No & Eczemas\\
    Psoriasis & No &\\
    \bottomrule
  \end{tabular}
  \label{tab:common_diseases}
\end{table}

This data will then act as a guide for what skin disease images to gather. A bulk image downloader is a good tool that can be used to achieve this.

\subsection{Machine Learning}

Once we have collected the data, the next step will be to prepare the data for the machine learning process. The process is depicted below: 

\begin{figure}[!ht]
  \centering
  \includegraphics[width=.6\textwidth]{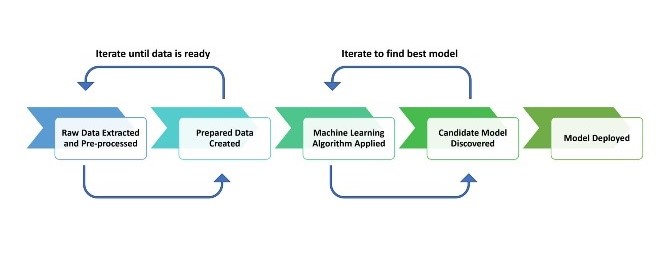}
  \caption{The Machine Learning Process}
  \label{fig:machine_learning}
\end{figure}

\newpage
\section{Results and Discussion}
\label{sec:resuts_and_discussion}

The machine learning approach showed very poor scores below 50\% accuracy as shown below:

\begin{figure}[!ht]
  \centering
  \includegraphics[width=.6\textwidth]{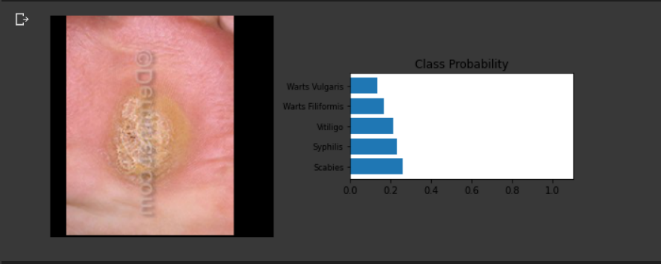}
  \caption{Machine Learning Evaluation Results}
  \label{fig:ml_evaluation_results}
\end{figure}

We also tried to reduce the number of diseases to train so we can understand the correlations between the different diseases. It appeared, that most images were indistinguishable.

\begin{figure}[!ht]
  \centering
  \includegraphics[width=.6\textwidth]{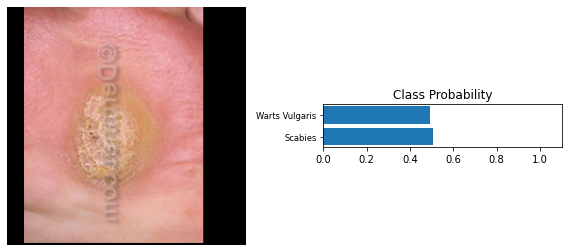}
  \caption{Machine Learning Evaluation Results for two classes}
  \label{fig:ml_2_classes}
\end{figure}

Further training did not also seem to show improvement as the running loss kept changing haphazardly

\begin{figure}[!ht]
  \centering
  \includegraphics[width=.6\textwidth]{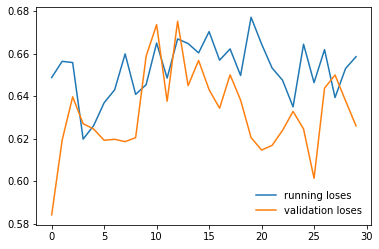}
  \caption{Training Performance}
  \label{fig:training_performance}
\end{figure}

A confusion matrix of two diseases that seemed to have similar characteristics was as below

\begin{figure}[!ht]
  \centering
  \includegraphics[width=.6\textwidth]{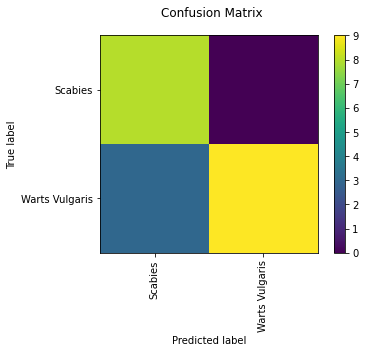}
  \caption{A Confusion Matrix for two classes}
  \label{fig:confusion_matrix}
\end{figure}

At this point, it was clear machine learning alone was not going to provide a solution so we tried a supervised learning approach.

\subsection{Object Detection}

In object detection, we are required to perform an additional data pre-processing step where we label the images and objects of interest on each image. A useful tool to do this is labellmg \cite{18}.

\begin{figure}[!ht]
  \centering
  \includegraphics[width=.6\textwidth]{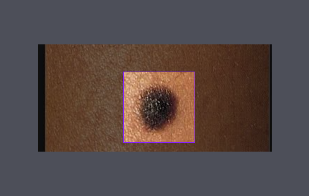}
  \caption{A labelled image of Kaposi sarcoma}
  \label{fig:labelled_kaposi}
\end{figure}

Once the labelling is done, the labelled images may be converted to different formats for processing. For this, roboflow provides a wide range of useful features. In our case, we need to convert to tf-records which is a TensorFlow data format that allows for the training with object detection.

\begin{figure}[!ht]
  \centering
  \includegraphics[width=.4\textwidth]{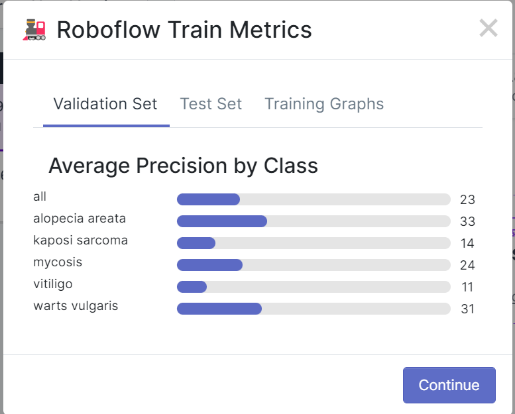}
  \caption{Average precision using roboflow}
  \label{fig:precision_roboflow}
\end{figure}

Once the images were labelled, we identified a suitable model to perform the training. Along the options were SSD, R-CNN and YOLO. Pulkit Sharma provides a good introduction to the various object detection models \cite{18}.

\subsubsection{R-CNN}

Region-based Convolutional Neural Networks are a family of neural networks that uses selective search to extract boxes from an image (these boxes are called regions). Again, Pulkit Sharma provides a more detailed overview of the working concepts of R-CNN.
R-CCN has a higher accuracy despite a slower training process. YOLO is faster with training but not as accurate as R-CNN.

\subsubsection{Colab}

Colab, or "Collaboratory", allows you to write and execute Python in your browser. It also provides access to a GPU. The training was then done using Colab by uploading a jupyter notebook, which is where we put all the preliminary python code ready for execution. With this, we were able to achieve accuracies higher than 60\%.

\begin{figure}[!ht]
  \centering
  \includegraphics[width=.4\textwidth]{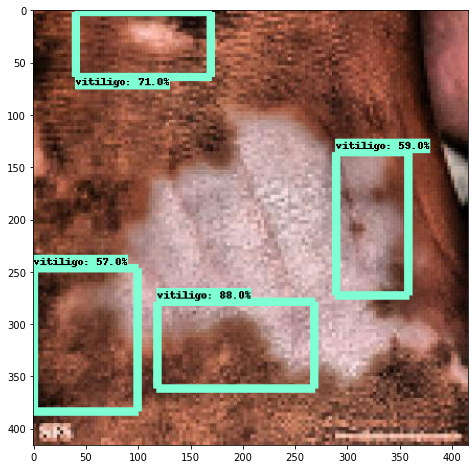}
  \caption{Results for Vitiligo }
  \label{fig:results_vitiligo}
\end{figure}

\begin{figure}[!ht]
  \centering
  \includegraphics[width=.4\textwidth]{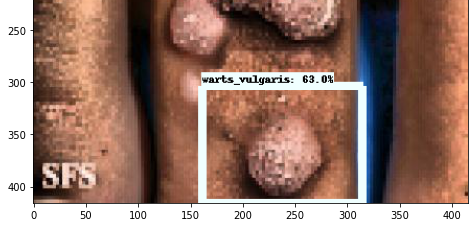}
  \caption{Results for Warts Vulgaris }
  \label{fig:results_vulgaris}
\end{figure}

\newpage
\section{Conclusion}
\label{sec:conclusion}

The results show us that skin disease diagnosis is very possible with image analysis. We were able to detect the diseases from the images with very high accuracies: 63\%, 88\% and 57\% which are all reasonable scores. When these scores are backed with historical data, we can more confidently diagnose patients. However, we do not provide much information on the historical processing. This will follow in a future research paper. We also restricted the number of classes/diseases to five only so that we could operate within the limits of colab training time.

\section{Future Scope}
\label{sec:future_scope}

Colab did not allow us to train for longer periods so using a computer with a dedicated GPU would greatly improve these results and allow us to train for longer periods.
With more labelled data, we can achieve high scores. More data may have to be collected from health centres and taking into account skin race may radically improve these scores.
Once these two recommendations are met, more diseases can be trained to allow the model to detect several other diseases.
A mobile application, as well as a web application, will then be constructed to integrate these techniques and make them available to skin patients over the internet.

%Bibliography
\bibliographystyle{unsrt}  
\bibliography{references}
\label{sec:references}

\end{document}